\documentclass{iopart}
\usepackage{amssymb}
\usepackage{indentfirst}
\usepackage{graphicx}
\usepackage{bm}

\begin{document}

\title[]
{Space-like and time-like pion-rho transition form factors in the
light-cone formalism}

\author{Jianghao Yu$^1$, Bo-Wen Xiao$^2$ and Bo-Qiang Ma$^{1,3}$}
\address{$^1$Department of Physics, Peking University, Beijing
100871, China}
\address{$^2$Department of Physics, Columbia University, New York,
NY 10027, USA}
\address{$^3$MOE Key Laboratory of Heavy Ion Physics,
Peking University, Beijing 100871, China}
\ead{mabq@th.phy.pku.edu.cn}
\begin{abstract}

Having calculated the light-cone wave function of the pseudoscalar
meson by using two equivalent and fully covariant methods, we
generalize such methods to the valence Fock states of the vector
meson in the light-cone formalism. We investigate the decay
constant of the $\rho$ meson $f_{\rho}$, the $\gamma^{\ast} \pi
\to \rho $ and $\gamma^* \rho \to \pi $ transition form factors
and especially the transition magnetic moments. By using two
groups of constraint parameters, we predict the space-like and
time-like form factors $F_{\pi \rho}(Q^2)$ and $F_{\rho \pi}(Q^2)$
at low and moderate energy scale and the electromagnetic radius of
these transition processes. In addition, we extend our calculation
to $\gamma^* \pi \to \omega$ space-like and time-like form factors
by using the same sets of parameters.

\end{abstract}

\maketitle

\section{Introduction}

To investigate the electromagnetic transition processes between
the pseudoscalar meson and the vector meson is an important and
distinctive way to understand the internal structure of hadrons.
It is well known that, at the low and moderately high four
momentum transfer $Q^2$, the elastic and transition form factors
of exclusive processes have to be treated non-perturbatively due
to the large coupling constant of QCD and bound state effects. In
non-perturbative region, the intrinsic momentum of quarks in a
meson has the same order of magnitude as that of the quark mass.
Therefore, the consistent treatment of relativistic effects of the
quark motion and spin in a bound state becomes a main issue. The
light-cone~(LC) formalism \cite{Dirac49,BPPrev98} provides a
convenient framework for the relativistic description of hadrons
in terms of quark and gluon degrees of freedom. Its application to
exclusive processes has mainly been developed in the light-cone
Fock representation \cite{BPPrev98,Fre06,LB80,BHMS01}.


Conventionally, the exclusive form factor can be calculated in terms
of the overlap of light-cone wave functions in the Drell-Yan frame
\cite{DrellYan70}, which refer to the valence contribution. However,
in many cases the non-valence section will contribute to the from
factor, even in the $q^+=0$ frame, which refer to the the zero-mode
contribution \cite{ChoiJi02,ChoiJi03,Jaus03}. Therefore, we need to
treat the form factor in the light-cone formulism carefully. In this
paper, we derived the spin-flavor section of valence light-cone
$\rho$ meson wave function through the Melosh transformation
\cite{melosh} of the instant SU(6) quark model. On the other hand,
from one kind of the covariant $\rho$-meson-quark-antiquark vertex
we also obtain this same spin-flavor section of the light-cone
$\rho$ meson wave function. In Ref.~\cite{Jaus03,Jaus91,ChoiJi03},
this vertex has been used to light-cone quark model for exclusive
processes. Bakker, Choi and Ji \cite{ChoiJi03} used this vector
meson vertex to calculate the transition form factors between
pseudoscalar and vector mesons in light-cone dynamics. They
concluded that the zero mode contribution vanished in the $q^+=0$
frame. Therefore, it is fortunate that we can discuss the valence
contribution only. For the momentum space section of light-cone wave
function, Bakker {\it et al.} used the perturbative energy
denominator to analyse the non-valence structure of the wave
function. But apparently hadronic wave function are
non-perturbative, so a broadly used Brodsky-Huang-Lepage wave
function \cite{BHL81} can reflect some feature of the
non-perturbative effect. In this paper, we specify the pseudoscalar
and vector meson to pion-rho transition process. Several papers has
discussed for this process. Choi and Ji~\cite{ChoiJi97} used a
different covariant $\rho$-quark-antiquark vertex~\cite{Ji92} to
discuss the space-like $\rho \to \pi \gamma^*$ process. Cardarelli
{\it et al.}~\cite{CarSim95} summed the Dirac and Pauli form factors
in the space-like $\gamma^* \pi^+ \to \rho^+ $ process, but Pauli
form factor perhaps has non-vanishing zero mode contribution and the
paper did not verify this. We will give an unified treatment for the
$\gamma^* \pi \to \rho $ process in the space-like and time-like
region and give a more careful description about $\rho$ meson decay
properties, transition magnetic moments and the electromagnetic
radius. Furthermore, other approaches have been adopted in order to
understand these transition form factors, such as the lattice
technique by Edwards~\cite{Edw05} for the $\gamma^* \rho \to \pi $
transition, the light-cone QCD sum rule calculation by
Khodjamirian~\cite{Kho99} for the $\gamma^* \rho \to \pi $
transition, and the model based on Dyson-Schwinger
equation~\cite{Maris02}.

The paper is organized as follow: we derive the wavefuntion of the
vector meson in Section \ref{sec2}. In Section \ref{sec3} and
Section \ref{sec4}, we compute the transition form factors and
decay widths, respectively. Finally, we provide numerical analysis
results of theoretical calculations. We present numerical analysis
and conclusion in Section \ref{sec5} and \ref{sec6}.

\section{The valence Fock state of the rho meson}\label{sec2}

The light-front Fock expansion of any hadronic system is
constructed by quantizing QCD at fixed light-cone time $x^+ = x^0
+ x^3$ and by establishing the invariant light-cone Hamiltonian
$H_{LC}$: $ H_{LC} = P^+ P^- - \vec{
P}_\perp^2$~\cite{Dirac49,BPPrev98}. In principle, solving the
$H_{LC}$ eigenvalue problem gives rise to the entire mass spectrum
of color-singlet hadron states in QCD, as well as their
corresponding light-front wave functions.  In particular, the
hadronic state satisfies $H_{LC} \vert \psi_H \rangle= M^2
\vert\psi_H \rangle$, where $\vert \psi_H \rangle$ is an expansion
in multi-particle Fock states. Considering a meson with momentum
$P$ and spin projection $S_z$, one can expand the hadronic
eigenstate $\vert\psi_M \rangle$ in QCD in terms of
eigenstates $\{\vert n \rangle \}$: 
\begin{eqnarray}
\fl \left\vert \psi_M (P^+, {\vec{P}_\perp}, S_z)\right> &=&
\sum_{n, \lambda_i\in n}\ \int \prod_{i=1}^{n} \left(  {{\rm
d}x_i\, {\rm d}^2 {\vec{k}_{\perp i}} \over  2 \sqrt{x_i}\, (2
\pi)^3}\ \right) \, 16\pi^3 \delta\left(1-\sum_{i=1}^{n}
x_i\right)\, \delta^{(2)}\left(\sum_{i=1}^{n} {\vec{k}_{\perp
i}}\right) \nonumber \\
&& \qquad \rule{0pt}{4.5ex} \times
\psi^{S_z}_{n/M}(x_i,{\vec{k}_{\perp i}}, \lambda_i) \left\vert
n;\, x_i P^+, x_i {\vec{P}_\perp} + {\vec{k}_{\perp i}},
\lambda_i\right>, \label{fockstate}
\end{eqnarray}
where the Fock state $n$ contains $n$ constituents and $\lambda_i$
is the helicity of the $i$-th constituent. Here $x_i = k^+_i/P^+$
are boost-invariant light-cone longitudinal momentum fractions and
${\vec{ k}_{\perp i}}$ represent the transverse momentum of the
$i$-th constituent in Fock state $n$ in the center of mass frame.
The $n$-particle Fock states are normalized as follows:
\begin{eqnarray}
\fl  \left< n;\, p'_i{}^+, {\vec{p\,'}_{\perp i}}, \lambda'_i
\right. \, \left\vert n;\, p^{~}_i{}^{\!\!+}, {\vec{p}_{\perp i}},
\lambda_i\right> = \prod_{i=1}^n \!\left( 16\pi^3
  p_i^+ \delta(p'_i{}^{+} - p^{~}_i{}^{\!\!+})\
 \! \delta^{(2)}( {\vec{p\,'}_{\perp i}} - {\vec{p}_{\perp i}})\
 \! \delta_{\lambda'_i \lambda^{~}_i}\!\right) .
\label{normalize}
\end{eqnarray}
To simplify the problem, we construct the light-cone wavefunctions
of mesons according to the valence quark model. Since the high
Fock states of hadron are suppressed \cite{LB80}, we only take the
valence states of the light-cone wave function into account, which
corresponds to the minimal Fock states or the first order
contributions in the full calculation.

For the pion meson, we have derived the valence light-cone
wavefunctions by employing two different methods~\cite{Xiao03}. In
the following, we summarize the main idea of two methods. On one
hand, one can write the vertex of the pseudoscalar (e.g., pion)
with two spin-$\frac{1}{2}$ fermions (e.g.,quark and antiquark)
inside as the following:
\begin{equation}
\frac{\overline{u}(p_{1}^{+},p_{1}^{-},\vec{k}_{\perp })}{\sqrt{
p_{1}^{+}}}\gamma _{5}\frac{v(p_{2}^{+},p_{2}^{-},-\vec{k}_{\perp
})}{\sqrt{p_{2}^{+}}},\label{pivertex}
\end{equation}
where $\bar{u}(p_{1}^{+},p_{1}^{-},\vec{k}_{\perp })$ and
$v(p_{2}^{+},p_{2}^{-},-\vec{k}_{\perp })$ are the light-cone
spinors of the quark and the antiquark, respectively. In this full
relativistic field theory treatment, this matrix element of the
interaction vertex~(\ref{pivertex}) can be taken as the minimal
Fock wavefunctions of pseudoscalars. On the other hand, the same
problem can be analyzed in the light-cone quark model. In this
model, the light-cone wave function of a composite system can be
obtained by transforming the ordinary equal-time (instant-form)
wave function in the rest frame into that in the light-front
dynamics, by taking into account the relativistic effects such as
the Melosh-Wigner rotation effect \cite{Ma93,Huang94}. The
transformation for the spin space wave functions between the two
formalisms is accomplished by the use of the Melosh-Wigner
rotation. The Melosh-Wigner rotation is one of the most important
ingredients of the light-cone formalism, and it relates the
light-cone spin state $|J,\lambda\rangle_F$ to the ordinary
instant-form spin state wave functions $|J,s\rangle_T$ by the
general relation~\cite{melosh,Ma91}
\begin{equation}
|J,\lambda\rangle_F=\sum_s
U^J_{s\lambda}|J,s\rangle_T.\label{meloshwigner}
\end{equation}
The transformation for the momentum space wave functions becomes
possible with the help of some Ansatz such as the
Brodsky-Huang-Lepage prescription~\cite{BHL81}. The equivalence of
the two approaches has been demonstrated in Ref.~\cite{Xiao03},
where both approaches are used to derive the pion light-cone wave
function and lead to the same result.

Hereafter, in a similar way, we extend such idea to vector mesons
and derive the light-cone wave functions of the $\rho$ meson as an
example. In the end, we also provide similar calculation respect
to the vector meson $\omega$.

\subsection{Vector meson wave-functions from Melosh-Wigner rotation.}

Using the Melosh-Wigner rotation, we derive the light-cone wave
function for the quark-antiquark Fock state of the $\rho$ meson in
the light-cone quark model. The spin parts of the $\rho$ wave
function in the SU(6) quark-antiquark model in the instant form are
written as: for the $\rho^+$ meson
\begin{eqnarray}
\vert \rho^+_{\pm} \rangle &=& u^{\uparrow , \downarrow} {\bar
d}^{\uparrow ,\downarrow}
\nonumber\\
\vert \rho^+_0 \rangle &=& \frac{1}{\sqrt{2}}(u^{\uparrow } {\bar
d}^{\downarrow}+u^{ \downarrow} {\bar d}^{\uparrow });
\label{instform1}
\end{eqnarray}
$u \leftrightarrow d$ in (\ref{instform1}) corresponds to the
$\rho^-$ meson.
Applying the transformation in equation~(\ref{meloshwigner}) on both
sides of equation~(\ref{instform1}), we obtain the spin space wave
function of the $\rho^+$ in the light-front frame. The Melosh-Wigner
transformation relates the instant-form and light-front form spin
states with four-momentum $(k^0,\vec{k})$ as follows
\begin{eqnarray}
 \chi_{q}^\uparrow(T)&=&w_q[(k^++m)\chi_q^\uparrow(F)-k^R\chi_q^\downarrow(F)],\nonumber\\
 \chi_{q}^\downarrow(T)&=&w_q[(k^++m)\chi_q^\downarrow(F)+k^L\chi_q^\uparrow(F)],\label{melosh}
\end{eqnarray}
where $w_q=[2k^+(k^0+m)]^{-\frac{1}{2}}$, $k^{R,L}=k^1\pm
\textmd{i}k^2 $, and $k^+=k^0+k^3=x\mathcal{M}$, $m$ is the mass
of the quark, and
$\mathcal{M}^2=\frac{\vec{k}_\perp^2+m^2}{x(1-x)}$ is the
invariance mass of the composite state. Therefore the spin part of
the light-cone wave function of the $\rho^+$ reads
\begin{equation}
\psi ^{S_z}_{\rho+}(x,\vec{k}_{\perp })=\sum_{\lambda _{1},\lambda
_{2}}C_{S_z}^{F}(x, \vec{k}_{\perp },\lambda _{1},\lambda
_{2})\chi _{1}^{\lambda _{1}}(F)\chi _{2}^{\lambda
_{2}}(F),\nonumber\\
 \label{spinwave}
\end{equation}%
where $S_z$ and $\lambda$ are the spin projection of the $\rho^+$
meson and the helicity of the quark. For the rho meson $\rho^+_+$
with $S_z=+1$, the coefficients of the spin wave function are
\begin{eqnarray}
C^{F}_{+1}(x,\vec{k}_{\perp},\uparrow,\uparrow)&=&\omega^{-1}(m(\mathcal{M}+2m)+\vec{k}^{2}_{\perp});\nonumber \\
C^{F}_{+1}(x,\vec{k}_{\perp},\uparrow,\downarrow)&=&\omega^{-1}(x\mathcal{M}+m)k^R; \nonumber \\
C^{F}_{+1}(x,\vec{k}_{\perp},\downarrow,\uparrow)&=&-\omega^{-1}((1-x)\mathcal{M}+m)k^R; \nonumber\\
C^{F}_{+1}(x,\vec{k}_{\perp},\downarrow,\downarrow)&=&-\omega^{-1}(k^R)^2,
 \label{coeff1}
\end{eqnarray}
where $\omega=(\mathcal{M}+2m)\sqrt{\vec{k}^{2}_{\perp}+m^2}$. The
coefficients of the $\rho^+_0$ are
\begin{eqnarray}
C^{F}_{0}(x,\vec{k}_{\perp},\uparrow,\uparrow)&=&\omega^{-1}((1-x)\mathcal{M}-x\mathcal{M})k^L /\sqrt{2};\nonumber \\
C^{F}_{0}(x,\vec{k}_{\perp},\uparrow,\downarrow)&=&\omega^{-1}(m(\mathcal{M}+2m)+2\vec{k}^{2}_{\perp})/\sqrt{2}; \nonumber \\
C^{F}_{0}(x,\vec{k}_{\perp},\downarrow,\uparrow)&=&\omega^{-1}(m(\mathcal{M}+2m)+2\vec{k}^{2}_{\perp}) /\sqrt{2};\nonumber\\
C^{F}_{0}(x,\vec{k}_{\perp},\downarrow,\downarrow)&=&\omega^{-1}(x\mathcal{M}-(1-x)\mathcal{M})k^R
/\sqrt{2}.
 \label{coeff0}
\end{eqnarray}
And the coefficients of the $\rho^+_-$ are
\begin{eqnarray}
C^{F}_{-1}(x,\vec{k}_{\perp},\uparrow,\uparrow)&=&-\omega^{-1}(k^L)^2;\nonumber \\
C^{F}_{-1}(x,\vec{k}_{\perp},\uparrow,\downarrow)&=&\omega^{-1}((1-x)\mathcal{M}+m)k^L; \nonumber \\
C^{F}_{-1}(x,\vec{k}_{\perp},\downarrow,\uparrow)&=&-\omega^{-1}(x\mathcal{M}+m)k^L; \nonumber\\
C^{F}_{-1}(x,\vec{k}_{\perp},\downarrow,\downarrow)&=&\omega^{-1}(m(\mathcal{M}+2m)+\vec{k}^{2}_{\perp}).
 \label{coeff-}
\end{eqnarray}
All of the component coefficients $C_{S_z}^{F}(x,\vec{k}_{\perp
},\lambda _{1},\lambda _{2})$ satisfy the unitary relation,
\begin{equation}
\sum_{\lambda _{1},\lambda _{2}}C_{S_z}^{F}(x,\vec{k}_{\perp
},\lambda _{1},\lambda _{2})^{\ast }C_{S_z}^{F}(x,\vec{k}_{\perp
},\lambda _{1},\lambda _{2})=1. \label{norm}
\end{equation}
Therefore, the Fock state expansion of the valence wave function
for the $\rho^+$ can be expressed as
\begin{eqnarray}
\fl \left\vert \psi _{\rho^+}\left( P^{+},\vec{P}_{\perp
},S_z\right) \right\rangle &=&\int
\frac{\mathrm{d}^{2}\vec{k}_{\perp }\mathrm{d}x}{16{ \pi }^{3}}
\left[ \psi _{\rho^+ }^{S_z}(x,\vec{k}_{\perp },\uparrow ,\uparrow
)\left\vert xP^{+},\vec{k}_{\perp },\uparrow
,\uparrow \right\rangle \right.\nonumber\\
&&\left. +\psi _{\rho^+ }^{S_z}(x,\vec{k}_{\perp },\uparrow
,\downarrow )\left\vert xP^{+}, \vec{k}_{\perp },\uparrow
,\downarrow \right\rangle  +\psi _{\rho^+ }^{S_z}(x,\vec{k}_{\perp
},\downarrow ,\uparrow )\left\vert xP^{+},\vec{k}_{\perp
},\downarrow ,\uparrow \right\rangle
\right. \nonumber\\
&&\left.+\psi _{\rho^+ }^{S_z}(x, \vec{k}_{\perp },\downarrow
,\downarrow )\left\vert xP^{+},\vec{k} _{\perp },\downarrow
,\downarrow \right\rangle \right]. \label{wavefunction}
\end{eqnarray}
Here the Fock state projection $\psi_{\rho^+}^{S_z}$ connects to
the wavefunction in the light-cone quark model as follow
\begin{eqnarray}
\psi_{\rho^+ }^{S_z}(x, \vec{k}_{\perp },\lambda_1 ,\lambda_2
)=C^{F}_{S_z}(x,\vec{k}_{\perp},\lambda_1 ,\lambda_2)\varphi
_{\rho^+ }(x,\vec{k}_{\perp}), \label{fockstate1}
\end{eqnarray}
where $\varphi _{\rho^+ }(x,\vec{k}_{\perp})$ is the momentum wave
function in the light-cone formalism. In ref.\cite{Huang94}, Huang,
Ma and Shen analyzed several valence wave function. Here we employ
the Brodsky-Huang-Lepage (BHL) prescription~\cite{BHL81} which is
used in many light-cone quark models~\cite{ChoiJi97,ChoiJi01}.
This wave function is as follow
\begin{equation}
\varphi _{\rho^+ }(x,\vec{k}_{\perp})=A\exp
\left[-\frac{1}{8{\beta }^{2}}\frac{\vec{k}_{\perp
}^{2}+m^{2}}{x(1-x)}\right].\label{BHL}
\end{equation}
From above, one can find that the Fock expansion of the two
particle Fock-state for $\rho$ meson with definite $S_z$ has four
possible spin configurations, in which each configuration have
different $\lambda_1$ and $\lambda_2$ and satisfy the spin sum
rule: $S_{z}=\lambda_1+\lambda_2+l_{z}$. While the spin of each
constituent undergoes a Melosh-Wigner rotation, the spin
components for these constituents does not necessarily conserve
the naive spin sum $S_z=\lambda_1+\lambda_2$ since there are
higher helicity components~\cite{Ma93,Huang94,Ma95,Cao97}. Such
higher helicity components plays an important role to understand
the proton ``spin puzzle'' in the nucleon case \cite{Ma91,Ma96}.

\subsection{Vector meson wave-functions from relativistic field theory.}

To obtain the spin wave function of the vector meson, we adopt
another way which is based on the full relativistic field theory
treatment of the interaction vertex following with the idea in
\cite{LB80,BHMS01,Xiao03}. Considering the $\rho$ vertices
connecting to the two spin-$\frac{1}{2}$ fermions (valence quark
and anti-quark), we choose the momentums in the standard
light-cone intrinsic frame
\begin{equation}
\begin{array}{lll}
P & = & (P^{+},\frac{M_{\pi}^{2}}{P^{+}},\vec{0}_{\perp }), \\
k_{1} & = & (xP^{+},\frac{\vec{k}_{\perp }^{2}+m^{2}}{xP^{+}},
\vec{k}_{\perp }), \\
k_{2} & = & ((1-x)P^{+},\frac{\vec{k}_{\perp }^{2}+m^{2}}{
(1-x)P^{+}},-\vec{k}_{\perp }), \\
q & = & (0,\frac{Q^{2}}{P^{+}},\vec{q}_{\perp }),\\
 P^{\prime }
& = & (P^{ +},\frac{M_{\rho}^{2}+\vec{q}_{\perp }^{2}}{P^{ +}},
\vec{q}_{\perp }),
\end{array}
\end{equation}
where $P$ and $P^{\prime }$ are the momenta of the vector meson
before interaction and after interaction, respectively; $k_{1}$
and $k_{2}$ are the momenta of the quark and anti-quark,
respectively; $q$ is the momentum of the virtual photon. In
addition, the polarization vectors of the $\rho$ meson used in
this analysis are given by
\begin{eqnarray}{\label{pol_vec}}
 {\epsilon}^{\mu}_{\pm}
&=&(\epsilon^+_{\pm},\epsilon^-_{\pm},{\bf \epsilon}^\perp_{\pm})
=\left(0,\frac{\mp \sqrt{2}P^{R,L}_{\perp}}{P^+}, \frac{\mp
1}{\sqrt{2}},\frac{-i}{\sqrt{2}}\right),
\nonumber\\
 {\epsilon}^{\mu}_0 &=& (\epsilon^+_0,\epsilon^-_0,{\bf \epsilon}^\perp_0)=\frac{1}{M}\left(P^+,\frac{\vec{
P}^2_{\perp}-M^2}{P^+}, \vec{ P}_{\perp}\right).\label{vect}
\end{eqnarray}
The matrix element of the effective vertex which had used in
\cite{ChoiJi03,Jaus03,Jaus91} is
\begin{equation}
\bar{u}(k_{1}^{+},k_{1}^{-},\vec{k}_{\perp
},\lambda_1)\left(\gamma
-\frac{k_1-k_2}{\mathcal{M}+2m}\right)\cdot \epsilon
_{S_z}v(k_{2}^{+},k_{2}^{-},-\vec{k}_{\perp },\lambda_2).
\label{vertex}
\end{equation}
We can obtain the Fock expansion of the two-particle Fock state of
the $\rho$ meson
\begin{eqnarray}
\fl  \left\vert \psi _{\rho^+}\left( P^{+},\vec{P}_{\perp
},S_z\right) \right\rangle &=&\int
\frac{\mathrm{d}^{2}\vec{k}_{\perp }\mathrm{d}x}{16{ \pi
}^{3}}\varphi _{\rho^+ }(x,\vec{k}_{\perp
})\nonumber\\
&&\times \bar{u}(k_{1}^{+},k_{1}^{-},\vec{k}_{\perp
},\lambda_1)\left(\gamma
-\frac{k_1-k_2}{\mathcal{M}+2m}\right)\cdot \epsilon
_{S_z}v(k_{2}^{+},k_{2}^{-},-\vec{k}_{\perp },\lambda_2)
\nonumber\\&& \left\vert xP^{+},\vec{k}_{\perp
 },\lambda_1,\lambda_2 \right\rangle,\label{fock}
\end{eqnarray}
where $\varphi _{\rho^+ }(x,\vec{k}_{\perp })$ is the momentum
space wavefunction of the $\rho$ meson. According to the spinor
representations of $\bar u$ and $v$ in the Appendix of the
ref~\cite{BPPrev98}, we calculate the above matrix element and
obtain the following results:
\begin{equation}
\begin{array}{lll}
\overline{u}_{\uparrow }(\gamma
-\frac{k_1-k_2}{\mathcal{M}+2m})\cdot \epsilon_+ v_{\uparrow } &=&
-\frac{\sqrt{2}(m(\mathcal{M}+2m)+\vec{k}^{2}_{\perp})}{\sqrt{x(1-x)}(\mathcal{M}+2m)},
 \\
\overline{u}_{\uparrow }(\gamma
-\frac{k_1-k_2}{\mathcal{M}+2m})\cdot \epsilon_+ v_{\downarrow }
&=&
-\frac{\sqrt{2}(k_{1}^{+}+m)k^R}{\sqrt{x(1-x)}(\mathcal{M}+2m)},
  \\
\overline{u}_{\downarrow }(\gamma
-\frac{k_1-k_2}{\mathcal{M}+2m})\cdot \epsilon_+ v_{\uparrow } & =
&
 \frac{\sqrt{2}(k_{2}^{+}+m)k^R}{\sqrt{x(1-x)}(\mathcal{M}+2m)},  \\
\overline{u}_{\downarrow }(\gamma
-\frac{k_1-k_2}{\mathcal{M}+2m})\cdot \epsilon_+ v_{\downarrow } &
= & \frac{\sqrt{2}(k^R)^2}{\sqrt{x(1-x)}(\mathcal{M}+2m)},
\end{array}\label{array11}
\end{equation}
for the $\rho^+$ meson with spin projection $S_z=+1$,
\begin{equation}
\begin{array}{lll}
\overline{u}_{\uparrow }(\gamma
-\frac{k_1-k_2}{\mathcal{M}+2m})\cdot \epsilon_0 v_{\uparrow } &=&
-\frac{(k_2^+-k_1^+)k^L}{\sqrt{x(1-x)}(\mathcal{M}+2m)},
 \\
\overline{u}_{\uparrow }(\gamma
-\frac{k_1-k_2}{\mathcal{M}+2m})\cdot \epsilon_0 v_{\downarrow }
&=&
-\frac{m(\mathcal{M}+2m)+2\vec{k}^{2}_{\perp}}{\sqrt{x(1-x)}(\mathcal{M}+2m)},
  \\
\overline{u}_{\downarrow }(\gamma
-\frac{k_1-k_2}{\mathcal{M}+2m})\cdot \epsilon_0 v_{\uparrow } & =
&
- \frac{m(\mathcal{M}+2m)+2\vec{k}^{2}_{\perp}}{\sqrt{x(1-x)}(\mathcal{M}+2m)},  \\
\overline{u}_{\downarrow }(\gamma
-\frac{k_1-k_2}{\mathcal{M}+2m})\cdot \epsilon_0 v_{\downarrow } &
= & -\frac{(k_1^+-k_2^+)k^R}{\sqrt{x(1-x)}(\mathcal{M}+2m)},
\end{array}\label{array22}
\end{equation}
for the $\rho^+$ meson with spin projection $S_z=0$, and
\begin{equation}
\begin{array}{lll}
\overline{u}_{\uparrow }(\gamma
-\frac{k_1-k_2}{\mathcal{M}+2m})\cdot \epsilon_- v_{\uparrow } &=&
\frac{\sqrt{2}(k^L)^2}{\sqrt{x(1-x)}(\mathcal{M}+2m)},
 \\
\overline{u}_{\uparrow }(\gamma
-\frac{k_1-k_2}{\mathcal{M}+2m})\cdot \epsilon_- v_{\downarrow }
&=& -\frac{\sqrt{2}(k^+_2+m)k^L}{\sqrt{x(1-x)}(\mathcal{M}+2m)},
  \\
\overline{u}_{\downarrow }(\gamma
-\frac{k_1-k_2}{\mathcal{M}+2m})\cdot \epsilon_- v_{\uparrow } &=&
 \frac{\sqrt{2}(k^+_2+m)k^L}{\sqrt{x(1-x)}(\mathcal{M}+2m)},  \\
\overline{u}_{\downarrow }(\gamma
-\frac{k_1-k_2}{\mathcal{M}+2m})\cdot \epsilon_- v_{\downarrow } &
= &
-\frac{\sqrt{2}(m(\mathcal{M}+2m)+\vec{k}^{2}_{\perp})}{\sqrt{x(1-x)}(\mathcal{M}+2m)},
\end{array}\label{array33}
\end{equation}
for the $\rho^+$ meson with spin projection $S_z=-1$. After
normalizations of the above matrix elements, we find
\begin{eqnarray}
\fl  C^{F}_{S_z}(x,\vec{k}_{\perp},\lambda_1,\lambda_2)=
-\frac{1}{\sqrt{2}\mathcal{M}}\overline{u}(k_{1}^{+},k_{1}^{-},\vec{k}_{\perp
},\lambda_1)(\gamma -\frac{k_1-k_2}{\mathcal{M}+2m})\cdot \epsilon
_{S_z}v(k_{2}^{+},k_{2}^{-},-\vec{k}_{\perp
},\lambda_2).~~~~~\label{cf}
\end{eqnarray}
Therefore, these two methods give exactly the same Fock expansion
of the $\rho^+$ meson and hence are equivalent to each other.

Similarly, we can find the Fock expansions of light-cone wave
functions for other vector mesons by employing the same
descriptions.

\section{Pion-rho and rho-pion transition form factors}\label{sec3}

With the vanishing zero mode contribution in the
pseudoscalar-vector transition process \cite{ChoiJi03}, the
light-front Fock expansion provides a Lorentz-invariant
representation of matrix elements of the electromagnetic current
in terms of the overlap of light-front wave functions of hadrons.
In section~\ref{sec2}, the Fock state expansions of $\rho$ and
$\pi$ light-cone wave functions are explicitly obtained. In this
section, we discuss the pion-rho transition and rho-pion
transition form factors.

The transition form factors of $\gamma^* \pi \to \rho$ and
$\gamma^* \rho \to \pi$ are defined by
\begin{eqnarray}
\Gamma ^{\mu }_{S_z,0}(\gamma^* \pi \to \rho)&=&-ieF_{\gamma
^{\ast }\pi \to \rho }(Q^{2})\varepsilon ^{\mu \nu \rho \sigma
}\mathcal{P}_{\nu }\epsilon^*
_{S_z ,\rho }q_{\sigma },\nonumber\\
\Gamma ^{\mu }_{S_z,0}(\gamma^* \rho \to \pi)&=&-ieF_{\gamma ^*
\rho \to \pi }(Q^{2})\varepsilon ^{\mu \nu \rho \sigma
}\mathcal{P}_{\nu }\epsilon _{S_z ,\rho }q_{\sigma
},\label{gammatran}
\end{eqnarray}
in which $\epsilon_{S_z} $ denotes the polarization vector of the
rho meson and $\mathcal{P}=\frac{1}{2}(P_{\pi}+P_{\rho})$. Here
$q=(P_{\rho}-P_{\pi})$ is the four-momentum transfer of the
virtual photon, $ q^{2}=-Q^{2}=q^{+}q^{-}-\vec{q}_{\perp
}^{2}=-\vec{q}_{\perp }^{2}$. Within the light-cone formalism, it
is well known that the form factors can be uniquely determined by
the matrix element of $\Gamma ^{+ }$ component of the
electromagnetic current,
\begin{eqnarray}
\fl  \Gamma ^{+}_{S_z,0}(\gamma^* \pi \to \rho)&=&\left\langle \psi
_{\rho }\left( P^{ +}_{\rho},\vec{P}_{\perp \rho},S_z \right)
\right\vert J^{+}\left\vert \psi _{\pi }\left(
P^{+}_{\pi},\vec{P}_{\perp \pi},0 \right) \right\rangle
\nonumber\\
&&\times\delta ^{2}(\vec{P}_{\pi\perp}+
\vec{q}_\perp-\vec{P}_{\rho\perp})\delta (P^+_{\pi}+ q^+-P^+
_{\rho}), \nonumber\\
\fl  \Gamma ^{+}_{0,S_z}(\gamma^* \rho \to \pi)&=&\left\langle \psi
_{\pi }\left( P^{ +}_{\pi},\vec{P}_{\perp \pi},0 \right) \right\vert
J^{+}\left\vert \psi _{\rho }\left( P^{+}_{\rho},\vec{P}_{\perp
\rho},S_z \right) \right\rangle\nonumber\\
&&\times \delta ^{2}(\vec{P}_{\rho\perp}+
\vec{q}_\perp-\vec{P}_{\pi\perp})\delta (P^+_{\rho}+ q^+-P^+
_{\pi}).~~~~~~\label{formfa}
\end{eqnarray}

In the $Q^2 \to 0 $ limit, the $\gamma^* \pi \to \rho$ transition
form factor is the same as $\gamma^* \rho \to \pi$ transition form
factor, which is namely the magnetic moment of the pion-rho
transition
\begin{eqnarray}
\mu_{\pi\rho}=F_{\gamma^* \pi \to \rho}(0)=F_{\gamma^* \rho \to
\pi}(0).\label{mu}
\end{eqnarray}
Another important static electromagnetic property of pion-rho and
rho-pion transitions is the charge radius of the pion-rho
transition, which can be obtained via the $Q^2 \to 0$ limit of
pion-rho or rho-pion transition form factors as
\begin{eqnarray}
\langle r_{\pi\rho}^2 \rangle &=& -6\, \lim_{Q^2\rightarrow
0}\frac{\partial F_{\gamma^* \pi \to \rho}\left(Q^2\right)}
{\partial Q^2}.\label{kss}
\end{eqnarray}

In these calculations, the plus component of the local
electromagnetic current reads
\begin{eqnarray}
\frac{\bar{u}(k^{\prime},\lambda^{\prime})}{\sqrt{k^{\prime
+}}}\gamma^+
\frac{\bar{u}(k,\lambda)}{\sqrt{k^{+}}}=2\,\delta_{\lambda^{\prime},\lambda},
\end{eqnarray}
and we choose the Drell-Yan assignment~\cite{DrellYan70}:
\begin{eqnarray}
q&=&\left(q^+,q^-,\vec{q}_{\perp}\right) =
\left(0,\frac{-q^2}{P^+},
\vec{q}_{\perp}\right), \nonumber \\
P&=&\left(P^+,P^-,\vec{P}_{\perp}\right) =
\left(P^+,\frac{M^2}{P^+},
\vec{0}_{\perp}\right).\label{drell-yan}
\end{eqnarray}
In this $q^+=0$ frame, the zero mode section (quark-antiquark pair
creation graph) gives no contribution to transition form factors
\cite{ChoiJi03}. Thus, the matrix elements of space-like currents
can be expressed as overlaps of light-cone wave functions with the
same number of Fock constituents. In particular, for the
transition form factors, we have
\begin{eqnarray}\label{eqn:overlap}
\fl  \frac{\Gamma ^{+}_{S_z,0}(\gamma^* \pi \to
\rho)}{2P^+}&=&\sum_a\int\frac{d^2\vec{k}_\perp dx}{16\pi^3}
\sum_{j,\lambda_1,\lambda_2} e_j\psi_{\rho,a}^{S_z
*}\left(x_i,\vec{k}^\prime_{\perp i}, \lambda_1,\lambda_2\right)
\psi_{\pi,a}\left(x_i,\vec{k}_{\perp i},
\lambda_1,\lambda_2\right),\nonumber\\
\fl  \frac{\Gamma ^{+}_{0,S_z}(\gamma^* \rho \to
\pi)}{2P^+}&=&\sum_a\int\frac{d^2\vec{k}_\perp dx}{16\pi^3}
\sum_{j,\lambda_1,\lambda_2} e_j\psi_{\pi,a}^{
*}\left(x_i,\vec{k}^\prime_{\perp i}, \lambda_1,\lambda_2\right)
\psi_{\rho,a}^{S_z}\left(x_i,\vec{k}_{\perp i},
\lambda_1,\lambda_2\right),~~~~~\label{formfac22}
\end{eqnarray}
where $e_j$ is the charge of struck constituents and $\psi_a
\left(x_i,\vec{k}_{\perp i},\lambda_i\right)$ is the light-cone
Fock expansion wave function respectively. Here, for the final
state light-cone wave function, the relative momentum coordinates
are
\begin{equation}
\vec{k}^\prime_{\perp i}=\vec{k}_{\perp
i}+\left(1-x_i\right)\vec{q}_{\perp}\label{kk}
\end{equation}
for the struck quark and
\begin{equation}
\vec{k}^\prime_{\perp i}=\vec{k}_{\perp i}-x_i\vec{q}_{\perp}
\end{equation}
for each spectator. It can be manifestly shown that the matrix
element $\Gamma ^{\mu }_{0,0}$ is vanishing for any $Q^2$ value and
the matrix element $\Gamma ^{\mu }_{+1,0}$ is the same as $\Gamma
^{\mu }_{-1,0}$. Therefore, evaluating $\Gamma ^{+ }_{+1,0}$ by
using the wave functions in equation~(\ref{fockstate}), we obtain
the explicit expressions for transition form factors.

For the $\gamma^*\pi^+ \to \rho^+$ transition, we have calculated
the light-cone valence quark wave function of the $\rho^+$ and
$\pi^+$. In the transition process, when $u$ quark is the struck
quark, the relative momentum coordinates of the final state are
$\vec{k}^\prime_{\perp u}=\vec{k}_{\perp
}+\left(1-x\right)\vec{q}_{\perp}=\vec{k}^\prime_{\perp} $ for the
$u$ quark and $\vec{k}^\prime_{\perp \bar{d}}=-\vec{k}_{\perp
}-(1-x)\vec{q}_{\perp}=-\vec{k}^\prime_{\perp} $ for the spectator
$\bar{d}$. The form factor is
\begin{eqnarray}
F^{u}_{\gamma ^{\ast }\pi \to \rho }(Q^{2}) =\frac{\Gamma
^{+}_{+1,0}}{ -ie(\vec{\epsilon}^*_{\perp }\times \vec{q}_{\perp
})\mathcal{P}^{+}} =e_{u}I(m,Q^2)\label{pi+rho+},
\end{eqnarray}
where
\begin{eqnarray}
I(m,Q^2)&=&2\int_{0}^{1}\mathrm{d}x\int \frac{\mathrm{d
}^{2}\vec{k}_{\perp }}{16{\pi }^{3}} \varphi^*_{\rho
}(x,\vec{k}_{\perp }^{\prime })\varphi _{\pi
}(x,\vec{k}_{\perp })\nonumber\\
&&\times \frac{m(\mathcal{M'}+2m)(k^{'L
}-k^L)q^R+k^{'L}q^R({k'}^Lk^R-{k'}^Rk^L)}
{Q^2(\mathcal{M'}+2m)\sqrt{\vec{k'}^{2}_{\perp}+m^2}\sqrt{\vec{k}^{2}_{\perp}+m^2}}\nonumber\\
&=&2\int_{0}^{1}\mathrm{d}x\int \frac{\mathrm{d
}^{2}\vec{k}_{\perp }}{16{\pi }^{3}} \varphi^*_{\rho
}(x,\vec{k}_{\perp }^{\prime })\varphi _{\pi
}(x,\vec{k}_{\perp })\nonumber\\
&&\times
\frac{m(\mathcal{M'}+2m)(1-x)+2(1-x)\vec{k}_{\perp}^2\sin^2\alpha
}
{(\mathcal{M'}+2m)\sqrt{\vec{k'}^{2}_{\perp}+m^2}\sqrt{\vec{k}^{2}_{\perp}+m^2}},\label{form11}
\end{eqnarray}
and $\mathcal{M'}^2=\frac{\vec{k'}_{\perp}^2+m^2}{x(1-x)}$. Here,
$\alpha$ is the angle between $\vec{k}_{\perp}$ and
$\vec{q}_{\perp}$. Similarly, when $\bar{d}$ quark is the struck
quark, the relative momentum of the struck quark $\bar{d}$ is $
\vec{k}^\prime_{\perp \bar{d}}=-\vec{k}_{\perp }+x\vec{q}_{\perp}$
and that of the spectator $u$ is $ \vec{k}^\prime_{\perp
\bar{d}}=\vec{k}_{\perp }-x\vec{q}_{\perp}$. If we let $x
\leftrightarrow (1-x)$ and $\vec{q}_{\perp} \leftrightarrow
-\vec{q}_{\perp}$ in the above formulae (This can be done because
$x$ is the integrate variable and form factor is only dependent on
$Q^2$), we directly see
\begin{equation}
F_{\gamma ^{\ast }\pi \to \rho
}^{\bar{d}}(Q^{2})=-e_{\bar{d}}I(m,Q^2).\label{form22}
\end{equation}
Therefore, the pion-rho transition form factor is
\begin{equation}
F_{\gamma ^{\ast }\pi \to \rho }(Q^{2})=F_{\gamma ^{\ast }\pi \to
\rho }^{u}(Q^{2})+F_{\gamma ^{\ast }\pi \to \rho
}^{\bar{d}}(Q^{2})=(e_u-e_{\bar{d}})I(m,Q^2).\label{pirho}
\end{equation}

In a similar way, we can calculate the $\gamma^* \pi^0\to \rho^0$
transition. The SU(6) instant quark wave functions of the $\rho^0
$ meson are
\begin{eqnarray}
\vert \rho^0_{\pm} \rangle &=& \frac{1}{\sqrt{2}}(u^{\uparrow ,
\downarrow } {\bar u}^{\uparrow , \downarrow}-d^{\uparrow ,
\downarrow} {\bar d}^{\uparrow , \downarrow});
\nonumber\\
\vert \rho^0_0 \rangle &=& \frac{1}{2}(u^{\uparrow } {\bar
u}^{\downarrow}+u^{ \downarrow} {\bar u}^{\uparrow }-d^{\uparrow }
{\bar d}^{\downarrow}-d^{ \downarrow} {\bar d}^{\uparrow
}),\label{rhomeson}
\end{eqnarray}
and the wave function of the $\pi^0 $ meson is
\begin{eqnarray}
\vert \pi^0 \rangle &=& \frac{1}{2}(u^{\uparrow } {\bar
u}^{\downarrow}-u^{ \downarrow} {\bar u}^{\uparrow }-d^{\uparrow }
{\bar d}^{\downarrow}+d^{ \downarrow} {\bar d}^{\uparrow
}).\label{rho334}
\end{eqnarray}
After the Melosh rotation, the light-cone quark wave functions can
be obtained. So the form factor is
\begin{eqnarray}
\fl  F_{\gamma ^{\ast }\pi^0 \to \rho^0
}(Q^{2})&=&\frac{1}{2}\left( F_{\gamma ^{\ast }\pi \to \rho
}^{u}(Q^{2})+F_{\gamma ^{\ast }\pi \to \rho
}^{\overline{u}}(Q^{2})\right)+\frac{1}{2}\left( F_{\gamma ^{\ast
}\pi \to \rho }^{d}(Q^{2})+F_{\gamma ^{\ast }\pi \to \rho
}^{\overline{d}}(Q^{2})\right)\nonumber\\
&=&\frac{1}{2}(e_u - e_{\overline{u}}+e_d-
e_{\overline{d}})I(m,Q^2)\nonumber\\
&=&F_{\gamma ^{\ast }\pi^+ \to \rho^+ }(Q^{2}). \label{pi0rho0}
\end{eqnarray}
We can find that equation~(\ref{pi0rho0}) is the same as
Eq.~(\ref{pi+rho+}), because both the instant wave functions and the
light-cone wave functions preserve the isospin symmetry. Thus at
this level by assuming $m_u=m_d=m$, the $\rho^{\pm}$ and $\rho^0$
have identical $\pi\gamma$ radiative decays. In this paper, we only
consider the case of isospin symmetry as other model
calculations~\cite{CarSim95,ChoiJi97,Maris02}, and will not
distinguish the $\rho^\pm$ transitions with $\rho^0$ transitions in
the following. In the Drell-Yan frame, the analytic continuation
from space-like to time-like region only requires the change of
$\vec{q}_{\perp}$ to $i\vec{q}_{\perp}$ in the form factor
\cite{ChoiJi01}. Choi and Ji \cite{ChoiJi01} showed that the
analytic continuation in the $q^+=0$ frame exhibits the same
behavior as the direct analysis in $q^+\neq 0$ frame by counting the
non-valence contribution.

Furthermore, the rho-pion transition form factor is
\begin{eqnarray}
\fl F_{\gamma ^{\ast }\rho \to \pi }(Q^{2}) &=&\frac{\Gamma
^{+}_{0,+1}}{ -ie(\vec{\epsilon}_{\perp }\times
\vec{q}_{\perp })\mathcal{P}^{+}}  \nonumber\\
&=&2(e_{u}-e_{\overline{ d}})\int_{0}^{1}\mathrm{d}x\int
\frac{\mathrm{d }^{2}\vec{k}_{\perp }}{16{\pi }^{3}} \varphi
^*_{\pi }(x,\vec{k}_{\perp }^{\prime })\varphi _{\rho
}(x,\vec{k}_{\perp })\nonumber\\
&&\times \frac{m(\mathcal{M}+2m)(k'^{R
}-k^R)q^L+k^{R}q^L({k'}^Rk^L-{k'}^Lk^R)}
{Q^2(\mathcal{M}+2m)\sqrt{\vec{k}^{2}_{\perp}+m^2}\sqrt{\vec{k'}^{2}_{\perp}+m^2}}\nonumber\\
&=&2(e_{u}-e_{\overline{ d}})\int_{0}^{1}\mathrm{d}x\int
\frac{\mathrm{d }^{2}\vec{k}_{\perp }}{16{\pi }^{3}} \varphi
^*_{\pi }(x,\vec{k}_{\perp }^{\prime })\varphi _{\rho
}(x,\vec{k}_{\perp })\nonumber\\
&&\times
\frac{m(\mathcal{M}+2m)(1-x)+2(1-x)\vec{k}_{\perp}^2\sin^2\alpha}
{(\mathcal{M}+2m)\sqrt{\vec{k}^{2}_{\perp}+m^2}\sqrt{\vec{k'}^{2}_{\perp}+m^2}}.\label{rhopi}
\end{eqnarray}

\section{Decay constants of the rho meson}\label{sec4}

Beside the transition form factors, we also compute the
electromagnetic decay properties of the $\rho$ meson by employing
the light-cone wave functions of the $\rho$ meson. Since the decay
channels of the $\rho$ meson are well measured, the decays
constants can be used as the constraints to fix the parameters in
numerical calculations.

\begin{enumerate}

\item Considering $\rho \to \pi \gamma$ decay process, the decay
width $\Gamma(\rho \to \pi \gamma)$ has the following relationship
with $F_{\gamma ^*\pi \to \rho}(0)$ and $F_{\gamma ^*\rho \to
\pi}(0)$~\cite{Jaus91}
\begin{eqnarray}
\fl \mathit{\Gamma } (\rho \to \pi \gamma
)=\frac{\alpha(M_{\rho}^2-M_{\pi}^2)^3}{24 M_{\rho}^3}{\vert
F_{{\gamma}^*\pi \to  \rho}
(0)\vert}^2=\frac{\alpha(M_{\rho}^2-M_{\pi}^2)^3}{24
M_{\rho}^3}{\vert F_{{\gamma}^*\rho \to  \pi}
(0)\vert}^2.~~~~~\label{gamma}
\end{eqnarray}
So if we get the transition magnetic moment $\mu_{\pi\rho}$, we
can calculate the decay width of the $\rho \to \pi \gamma$
process.

\item Another decay channel is the leptonic decay of the rho
meson. The electromagnetic decay constant $f_{\rho}$ is defined
from $\rho^0 \to e^+ e^- $~\cite{Jaus91} decay process. The width
for this process is given, for leptons of zero mass, by
\begin{equation}
\Gamma(\rho^0 \to e^+ e^-)=\frac{4\pi \alpha^2
f_{\rho}^2}{3M_{\rho}}. \label{rhodecayee}
\end{equation}
The matrix element of the local vector current is defined as
\begin{equation}
< 0 \,\vert J^\mu_{V} \vert \psi_{\rho}^{S_z}(P^+,\vec{P}_{\perp}) >
=i M_{\rho} f_{\rm \rho} \epsilon^\mu_{s_z}.\label{veccurr}
\end{equation}
In light-cone formalism, the plus component of the matrix element
can be written as
\begin{eqnarray}
\fl  < 0\, |J^+_{V}|P,h
>
&=&\frac{(e_{u}-e_d)}{\sqrt{2}}\int_{0}^{1}\mathrm{d}x\int
\frac{\mathrm{d }^{2}\vec{k}_{\perp }}{16{\pi }^{3}}\,{\varphi
}_{\rho}(x,\vec{k}_{\perp })\nonumber\\
&&\times
\sum_{\lambda_1,\lambda_2}\overline{v}(k_{2}^{+},k_{2}^{-},-\vec{k}_{\perp
},\lambda_2)\gamma^+ u(k_{1}^{+},k_{1}^{-},\vec{k}_{\perp
},\lambda_1)
\nonumber\\
&&\overline{u}(k_{1}^{+},k_{1}^{-},\vec{k}_{\perp
},\lambda_1)(\gamma -\frac{k_1-k_2}{M+2m})\cdot \epsilon
(J^z)v(k_{2}^{+},k_{2}^{-},-\vec{k}_{\perp
},\lambda_2),~~~~~~\label{f1plus}
\end{eqnarray}
so we get
\begin{equation}
\frac{f_{\rho }}{2\sqrt{3}}=\int_{0}^{1}dx\int
\frac{d^{2}\vec{k}_{\perp }}{16{\pi }
^{3}}\frac{m(M+2m)+2\vec{k}_{\perp
}^{2}}{(M+2m)\sqrt{{m}^{2}+{{\vec{k}_{\perp }}^{2}}}}{\varphi
}_{\rho}(x,\vec{k}_{\perp }). \label{decayB}
\end{equation}

\item From the electromagnetic decay, we can also relate the
electromagnetic decay to the weak decay $\rho^+ \to e^++\nu_e$ and
$\rho^+ \to \mu^++\nu_{\mu}$ by the following relation:
\begin{equation}
\Gamma(\rho^+ \to e^++\nu_e)=\frac{\alpha}{3}
G_F^2M_{\rho}^3f_{\rho}^2={ G_F^2M_{\rho}^4\Gamma(\rho^0\to
e^+e^-)\over 4\pi\alpha}.\label{rhoee}
\end{equation}
\end{enumerate}

\section{Numerical results and discussions}\label{sec5}

In the formula for the transition form factors $F_{\gamma ^{\ast
}\pi \to \rho }(Q^{2})$ and $F_{\gamma ^{\ast }\rho \to \pi
}(Q^{2})$, there are five parameters: the $\pi$ and $\rho$
normalization constant $A_{\pi}$ and $A_{\rho}$, the $\pi$
harmonic scale of the momentum space wave function $\beta_{\pi}$,
the $\rho$ harmonic scale $\beta _{\rho}$, and the quark masses
$m$ (assuming $m_u=m_d=m$ according to isospin symmetry). These
parameters can be fixed by five constraints and hence we can
proceed to calculate the transition form factors numerically, and
predict other transition properties. We list these constraints as
follows.

\begin{enumerate}
\item The weak decay constant $f_{\pi}=(92.4\pm0.25)~\mbox{MeV}$
~\cite{Caso98} defined from $\pi\to \mu\nu$ decay process by
$<0\,|\overline{u}
\gamma^{+}(1-\gamma_{5})d|\pi>=-\sqrt{2}f_{\pi}p^{+}$, thus one
obtains~\cite{Ma93,Xiao03}
\begin{equation}
\frac{f_{\pi }}{2\sqrt{3}}=\int_{0}^{1}dx\int
\frac{d^{2}\vec{k}_{\perp }}{16{\pi }
^{3}}\frac{m}{\sqrt{{m}^{2}+{{\vec{k}_{\perp }}^{2}}}}{\varphi
}_{\pi}(x,\vec{k}_{\perp }). \label{decayB1}
\end{equation}

\item The decay width $\mathit{\Gamma }(\pi ^{0}\to \gamma \gamma
)$ has the following relation with $F_{\pi \gamma }(0)\ $and $
F_{\gamma\pi }(0)$ ~\cite{CLEO98}:
\begin{equation}
\left\vert F_{\gamma \gamma ^{\ast }\to \pi }(0)\right\vert
^{2}=\left\vert F_{\gamma^*\pi \to \gamma  }(0)\right\vert
^{2}=\frac{64\pi \mathit{\Gamma } (\pi ^{0}\to \gamma \gamma
)}{(4\pi \alpha )^{2}M_{\pi }^{3}}.\label{with}
\end{equation}
We use $\mathit{\Gamma }(\pi ^{0}\to \gamma \gamma )=(7.74\pm
0.54)\mbox{ eV}$ \cite{CLEO98}, which leads to $F_{\pi \gamma
}(0)=(0.27\pm 0.01)\mbox{ GeV}^{-1}$ in our calculation.

\item The charge form factor of the pion has the following
low-energy expansion:
\begin{equation}
F_{\pi^+}(Q^2)=1+\frac{1}{6}{\langle r^2 _{\pi^+} \rangle}
Q^2+\mathcal{O}(Q^4),\label{ff}
\end{equation}
where $\langle r^2 _{\pi^+} \rangle$ is the electromagnetic radius
of the charged pion. Thus,
\begin{equation}
\langle r_{\pi ^{+}}^{2}\rangle =-6\frac{\partial F_{\pi
^{+}}(Q^{2})}{
\partial Q^{2}}|_{Q^{2}=0}.\label{rpirho}
\end{equation}
Experimentally, one finds~\cite{Dally82}
\begin{equation}
\langle r^2_{\pi^+}\rangle_{\mbox{exp}}=(0.439\pm 0.03)\mbox{
fm}^{2}.\label{exp}
\end{equation}

\item The value of the transition form factor at $Q^2=0$ (the
so-called transition magnetic moment) have been experimentally
determined from the radiative decay width of the $\rho$ meson. The
decay width of $\rho^\pm \to \pi^\pm \gamma$ process is given by
\begin{eqnarray}
\fl  \mathit{\Gamma } (\rho^\pm \to \pi^\pm \gamma
)=\frac{\alpha(M_{\rho}^2-M_{\pi}^2)^3}{24 M_{\rho}^3}{\vert
F_{{\gamma}^*\pi^{\pm} \to \rho^{\pm}}
(0)\vert}^2=\frac{\alpha(M_{\rho}^2-M_{\pi}^2)^3}{24
M_{\rho}^3}{\vert \mu_{\pi \rho} \vert}^2,~~~~~~~\label{decayrho}
\end{eqnarray}
and the experimental value~\cite{PDG04} is $\mathit{\Gamma
}(\rho^\pm \to \pi^\pm \gamma )=(68\pm 7)\mbox{ keV}$. Hence, the
value of the transition magnetic moment is
\begin{equation}
\mu_{\pi \rho}=F_{ \gamma^*\pi \to \rho}(0)=(0.733\pm 0.038
)\,\mbox{GeV}^{-1}.\label{pr}
\end{equation}

\item The experimental value~\cite{PDG04} of the decay width for
the $\rho^0 \to e^+ e^- $ process is $\Gamma(\rho^0 \to e^+
e^-)=(7.02 \pm 0.11)\mbox{keV}$. From the
Equation~(\ref{rhodecayee}), the rho decay constant $f_{\rho}$ can
be determined from this process. The resulting rho dacay constant
is
\begin{equation}
f_{\rho}=(155.6 \pm 19.5 )~\mbox{MeV}.\label{frho}
\end{equation}

\end{enumerate}

\begin{figure}
\centering\scalebox{1.0}{\includegraphics{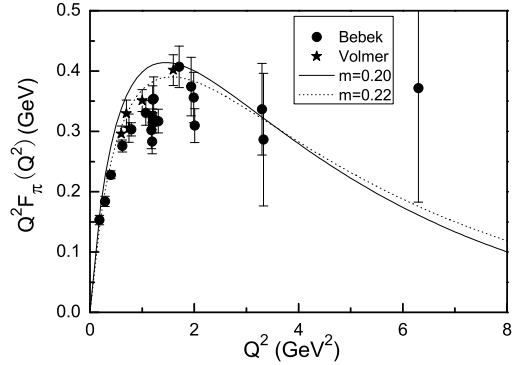}}
\caption{\label{fig:pionqff}The charge form factor of the pion
multiplied by $Q^2$ in few GeV of the  the momentum transfer. The
experimental data are taken from \cite{Bebek78} (full circles) and
\cite{Volmer01} (stars). The solid and dotted curves respectively
correspond to the results of the first set and the other two sets
of parameters. }
\end{figure}

\begin{figure}
\centering\scalebox{1.0}{\includegraphics{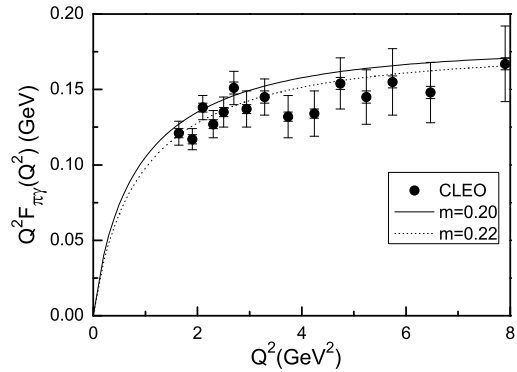}}
\caption{\label{fig:piongammaqff}The photon-pion transition form
factor multiplied by $Q^2$ compared with the experimental data.
The data (full circles) are taken from \cite{CLEO98}. The solid
and dotted curves respectively correspond to the results of the
first set and the other two sets of parameters.}
\end{figure}

From above five constraints, we can obtain the values of
parameters. For the parameters of the pion meson, we still adopt
the values of Ref.~\cite{Xiao03}, $m=0.200$~GeV,
$\beta_{\pi}=0.410$~GeV, and $A_{\pi}=47.5 ~\mbox{MeV}^{-1}$. For
the other parameters about $\rho$ meson, the values are
$\beta_{\rho}=0.410$~GeV and $A_{\rho}=39 ~\mbox{MeV}^{-1}$. This
is listed in the Set~I of table~\ref{tab:parameters}. Conversely,
using the above parameters, we can compute and predict the static
properties and transition form factors. The results of static
properties are shown in the Set~I of table~\ref{tab:properties}
and they are in very good agreement with above five experimental
constraints. Furthermore, as shown in figure~\ref{fig:pionqff} and
Figure~\ref{fig:piongammaqff}, these fixed parameters provide
predictions which are in good accordance with the experimental
data for the pion charge form factors and pion-photon transition
form factor, respectively. In addition, we can predict that the
electromagnetic radii of form factors is $\langle r_{\pi \rho
}^{2}\rangle =0.299~\mbox{fm}^2$.

\begin{table}
\caption{\label{tab:parameters}The three sets of parameters used
in the calculations.}
\begin{tabular}{@{}llll}
\br
 & Set I & Set II & Set III  \\
\mr
$m_q$ (GeV) & 0.200 & 0.220 & 0.220  \\
$\beta_{\pi}$ (GeV) & 0.410 & 0.420 & 0.420  \\
$\beta_{\rho}$ (GeV) & 0.410 & 0.420 & 0.360  \\
\br
\end{tabular}
\end{table}

\begin{table}
\caption{\label{tab:properties}The static electromagnetic
properties of pion and rho for the three sets of parameters.}
\begin{tabular}{@{}lllll}
\br
        & Set I & Set II & Set III & Experimental \\
\mr
$f_{\pi}(\rm MeV)$ & 92.4 & 92.0 & 92.0  & 92.4 \\
$\langle r^2_{\pi^+}\rangle (\rm fm^2) $ &0.446 & 0.351 & 0.351  & 0.439 \\
$F_{\pi \to \gamma \gamma}(\rm GeV^{-1})$ & 0.271 & 0.235 & 0.235 & 0.27 \\
$f_{\rho}(\rm MeV)$ & 155.9 & 181.2 & 155.9  & 155.6 \\
$F_{\rho \to \pi \gamma}(\rm GeV^{-1}) $ & 0.731 & 0.736 & 0.769  & 0.733 \\
$\langle r^2_{\pi\rho}\rangle(\rm fm^2)$ & 0.299 & 0.270 & 0.306  & \mbox{no Exp.} \\
\br
\end{tabular}
\end{table}

In above parametrization scheme, we adopt five experimental
constraints to fix the parameters. In an alternative way, the
normalizations of both $\rho$ and $\pi$ wave functions can be used
as one kind of theoretical constraints. We use the value of the
relativistic quark mass $m=0.22~\mbox{GeV}$ which is frequently
adopted. In set~II of table~\ref{tab:parameters}, we simple use the
same harmonic scales of the momentum space and in set~III, the
typical harmonic scale $\beta=0.360~\mbox{GeV}$ is adopted to the
$\rho$ meson. In order to fix the harmonic scale of the pion meson
we have to employ one experimental input: $f_{\pi}$. From the above
theoretical constraints and the only experimental value of
$f_{\pi}$, we can predict all other static properties which are
listed in set~II and set~III of the table~\ref{tab:properties}.
Also, the pion charge form factors and pion-photon transition form
factor from the predictions by using these parameters are in a very
good agreement with the experimental data, as shown in
figure~\ref{fig:pionqff} and figure~\ref{fig:piongammaqff}.

\begin{figure}
\centering\scalebox{1.0}{\includegraphics{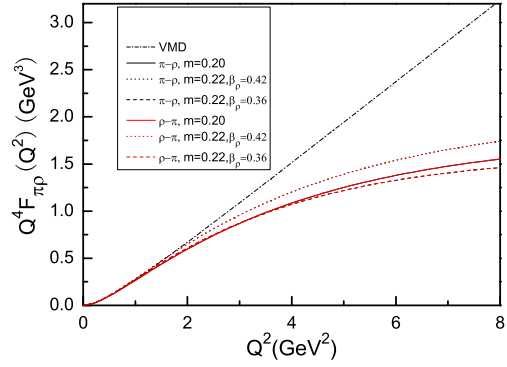}}
\caption{\label{fig:pionrhorhopionqff} The space-like pion-rho and
rho-pion transition for the $Q^4F_{\pi\rho}(Q^2)$ and
$Q^4F_{\rho\pi}(Q^2)$. The solid, dotted, and dashed curves
respectively correspond to the pion-rho form factors by using the
parameters of three sets of table~\ref{tab:parameters}, and the
other curves correspond to the rho-pion form factors. Furthermore,
the dash-dotted curve is the prediction of the $\rho$-pole vector
meson domination (VMD) model.}
\end{figure}

Figure~\ref{fig:pionrhorhopionqff} predicts the theoretical values
of pion-rho and rho-pion space-like transition form factors in a
range from low $Q^2$ to $Q^2=8~\mbox{GeV}^2$. Furthermore, we
compare our results with the curve of the $\rho$-pole vector meson
domination (VMD) model, where
$F^{VMD}_{\pi\rho}(Q^2)=F_{\pi\rho}(0)/(1+Q^2/M_{\rho}^2)$.
Although $F_{\rho \gamma ^{\ast }\to \pi }(Q^{2})$ and $F_{\gamma
^{\ast }\pi \to \rho }(Q^{2})$ are physically different in
equations~(\ref{rhopi}) and (\ref{pirho}),
figure~\ref{fig:pionrhorhopionqff} indicates that they are
numerically identical at very high precision, which implies that
$F_{\rho \gamma ^{\ast } \to \pi }(Q^{2})$ and $F_{\gamma ^{\ast
}\pi \to \rho }(Q^{2})$ have the same $Q^{2}$ dependence. The
identification of equations~(\ref{pirho}) and (\ref{rhopi}) can be
proven by making the variable transformation: $\vec{k}_{\perp }
\to \vec{k}_{\perp }-(1-x) \vec{q}_{\perp }$ and then
$\vec{q}_{\perp } \to -\vec{q}_{\perp }$ for
equation~(\ref{rhopi}).

We can extend our discussion to the $\gamma^* \pi^0\to \omega^0$
transition. If we ignore the mixing of $\omega$ and $\phi$, there
are only the $\gamma^*\pi \to \omega$ transition. The instant wave
function of the $\omega^0 $ meson are
\begin{eqnarray}
\vert \omega^0_{\pm} \rangle &=& \frac{1}{\sqrt{2}}(u^{\uparrow ,
\downarrow } {\bar u}^{\uparrow , \downarrow}+d^{\uparrow ,
\downarrow} {\bar d}^{\uparrow , \downarrow});
\nonumber\\
\vert \omega^0_0 \rangle &=& \frac{1}{2}(u^{\uparrow } {\bar
u}^{\downarrow}+u^{ \downarrow} {\bar u}^{\uparrow }+d^{\uparrow }
{\bar d}^{\downarrow}+d^{ \downarrow} {\bar d}^{\uparrow
}).\label{omega0}
\end{eqnarray}
For simplicity, we use the same parameters as the $\rho$ meson and
hence the SU(6) symmetry has not been violated. The form factor
can be written as
\begin{eqnarray}
\fl F_{\gamma ^{\ast }\pi^0 \to \omega^0
}(Q^{2})&=&\frac{1}{2}\left(F_{\gamma ^{\ast }\pi \to \rho
}^{u}(Q^{2})+F_{\gamma ^{\ast }\pi \to \rho
}^{\overline{u}}(Q^{2})\right)-\frac{1}{2}\left(F_{\gamma ^{\ast
}\pi \to \rho }^{d}(Q^{2})+F_{\gamma ^{\ast }\pi \to \rho
}^{\overline{d}}(Q^{2})\right)\nonumber\\
&=&\frac{1}{2}(e_u - e_{\overline{u}}-e_d+
e_{\overline{d}})I(m,Q^2)\nonumber\\
&=&3 F_{\gamma ^{\ast }\pi^+ \to \rho^+ }(Q^{2}).\label{omegaf}
\end{eqnarray}
Again, this result is derived from the isospin symmetry. Therefore,
if we adopt the first kind of parametrizations in table
\ref{tab:parameters}, we get the following predictions for the decay
width of the $\omega \to \pi \gamma$ transition and electromagnetic
radius:
\begin{eqnarray}
\mathit{\Gamma } (\omega\to\pi\gamma
)&=&\frac{\alpha(M_{\omega}^2-M_{\pi}^2)^3}{24 M_{\omega}^3}{\vert
F_{\omega\to\pi\gamma}
(0)\vert}^2=(642\pm 66)~\mbox{keV},\nonumber\\
\langle
r^2_{\pi\omega}\rangle&=&0.897~\mbox{fm}^{2}.\label{omegag}
\end{eqnarray}
and the experimental value of the decay width is
$\Gamma^{\mbox{exp}}(\omega\to\pi\gamma)=(717\pm51)~\mbox{keV}$.
Furthermore, the transition form factor $F_{\gamma^* \pi \to
\omega}(Q^2)$ is three times larger than the form factor of the
$\gamma^*\pi \to \rho$ process.

\begin{figure}
\centering\scalebox{1.0}{\includegraphics{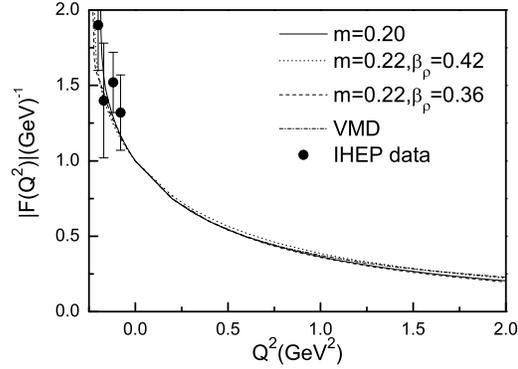}}
\caption{\label{fig:pionrhoff}Theoretical prediction for
$\gamma^*\pi \to \rho$ and $\gamma^*\pi \to \omega$ space-like and
time-like form factors, which have been normalized. Because of
isospin symmetry, these two form factors are the same curve. The
solid, dotted, and dashed curves respectively correspond to the form
factors by using the parameters of three sets of
table~\ref{tab:parameters}, and the dash-dotted curve is the
prediction of the $\rho$-pole vector meson domination (VMD) model.
The data in the time-like region are from \cite{Maris02,Vik81}}
\end{figure}

\begin{figure}
\centering\scalebox{1.0}{\includegraphics{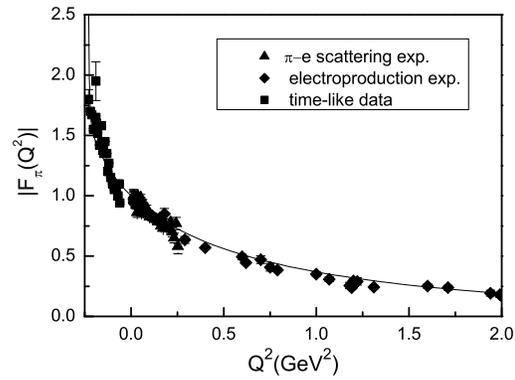}}
\caption{\label{fig:piontimeff}Pion space-like and time-like form
factors as a function of the momentum transfer $Q^2$. The data in
the spacelike region are from $\pi$-$e$ scattering experiments
\cite{pislff_sc1,pislff_sc2,pislff_sc3,pislff_sc4} and
electroproduction experiments
\cite{pislff_elprod1,pislff_elprod1,pislff_elprod2,pislff_elprod3,Bebek78,Volmer01}.
The data in the time-like region is from \cite{baldini}. }
\end{figure}

In figure~\ref{fig:pionrhoff}, the space-like and time-like
transition form factors are unified by analytical continuation
technique. For simplicity, we normalize the $\gamma^*\pi \to \rho$
and $\gamma^*\pi \to \omega$ form factors in $Q^2=0$ and so these
two form factors become the same curve in the
Figure~\ref{fig:pionrhoff}. In the time-like region, beyond the
threshold momentum transfer $Q^2$ ($Q^2 \simeq -0.2~
\mbox{GeV}^2$) the singularity for bound state production and the
imagine part will appear. Choi and Ji~\cite{ChoiJi01} treated this
case systematically for other processes. Now we only give the
results below the particle production threshold. And in time-like
region we can compare our numerical values with
data~\cite{Maris02,Vik81}. Besides, the same analytical
continuation technique can also be used to the time-like pion form
factor below threshold of particle production. In
Figure~\ref{fig:piontimeff}, we only show the theoretical
prediction by using the second set of the pion meson parameters in
Table~\ref{tab:parameters} and compare the results with the data
of many experiments.

\section{Conclusion}\label{sec6}

The light-cone formulism provides a convenient framework for the
relativistic description of hadrons in terms of quark and gluon
degrees of freedom, especially for the application to exclusive
processes. To obtain the light-cone wave functions of the $\rho$
meson, we employ both the light-cone quark model and
rho-quark-antiquark vertex description which are equivalent with
each other in the valence Fock-state expansion. Because of the
vanishing zero mode by using this vertex, it is easy to derive the
formulae of pion-rho transition form factors. Then we point out that
in the isospin symmetry limit the form factor of the
$\gamma^*\pi^\pm \to \rho^\pm$ transition is equal to that of the
$\gamma^*\pi^0\to \rho^0$ transition. We calculate the numerical
results of space-like and time-like pion-rho and rho-pion transition
form factors in terms of two groups of constraint parameters. The
first one is that we adopt five experimental constraints to fix five
parameters and conversely predict the static properties and
transition form factors. At the same time, we explore the
electro-weak decay properties of $\rho$ and $\pi$ mesons in
light-cone formulism to connect the parameters with experimental
constraints. Another parametrization is that we employ four
theoretical constraints and only one experimental input $f_\pi$ to
reduce the dependence of experimental values and enhance the ability
of prediction of this model. Finally, according to the isospin
symmetry, we extend our calculation to the space-like and time-like
$\gamma^*\pi\to\omega$ processes and compare the theoretical result
with time-like data.


\ack J.Y. thank Prof.\,Chueng-Ryong Ji for his encouragement and
discussion about zero mode. This work is partially supported by
National Natural Science Foundation of China (Nos.~10421503,
10575003, 10528510), by the Key Grant Project of Chinese Ministry
of Education (No.~305001), and by the Research Fund for the
Doctoral Program of Higher Education (China).

\def\Journal#1#2#3#4{#4 {#1} {\bf#2} #3}
\def\NCA{\it Nuovo Cimento}
\def\NP{\it Nucl. Phys.}
\def\NPA{{\it Nucl. Phys.} A}
\def\NPB{{\it Nucl. Phys.} B}
\def\NPBP{{\it Nucl. Phys.}B (Proc.Suppl.)}
\def\PLB{{\it Phys. Lett.}  B}
\def\PRt{\it Phys. Rep.}
\def\PRL{\it Phys. Rev. Lett.}
\def\PR{\it Phys. Rev.}
\def\PRD{{\it Phys. Rev.} D}
\def\PRC{{\it Phys. Rev.} C}
\def\RMP{\it Rev. Mod. Phys.}
\def\ZPA{{\it Z. Phys.} A}
\def\ZPC{{\it Z. Phys.} C}
\def\AM{{\it Ann. Math.}}
\def\JPG{\it J. Phys. G: Nucl. Part. Phys. }
\def\EPJA{\it Eur. Phys. J. A }
\def\EPJC{\it Eur. Phys. J. C }

\end{document}